\documentclass[11pt, oneside]{article}

\usepackage{caption}
\usepackage{graphicx}
\usepackage{bm}
\usepackage{geometry} 

\title{On the Penrose Process for Rotating Black Holes}
\author{Leon Heller
\\Biological and Quantum Physics Group, P-21, Los Alamos 
National Laboratory \\ Los Alamos, New Mexico, 87544}

\begin{document}

\maketitle

\begin{abstract}

Penrose described a process that, in principle, could extract energy and angular momentum from a rotating black hole.  Here we examine  two  procedures that were claimed to be capable of implementing  the Penrose idea;  both make use of a particle  moving at the horizon. In one, the particle is swallowed, and in the other the particle and black hole gradually exchange energy and angular momentum. We show that if the particle has negative energy and negative angular momentum but no radial momentum both procedures violate the requirement that the area of a black hole not decrease. For the gradual exchange method, however,  it appears that the Penrose process could proceed if the particle has positive energy and angular momentum, but nevertheless removes energy from the black hole. It does not, however, lead to a Schwarzschild black hole. For an extreme Kerr black hole it's mass decreases by at most 9.7\%, well short of the theoretical limit for a reversible process of $1-1/\sqrt{2} =29\%$.

\end{abstract}

\maketitle

\section{Introduction} 

The  Penrose proprosal for extracting energy from a rotating black hole \cite{Penrose} is based on the  possibility of  introducing into the ergosphere, the region between the horizon and  the static limit, a particle having negative energy  as measured at infinity. For example, a positive energy particle decays into two, one of which  has sufficiently large angular momentum   opposite to the rotation of the black hole. If the black hole were able to take up some or all of this particle's negative energy and angular momentum, the corresponding amount of the black hole's positive energy would have been transferred to the other particle resulting from the decay, which can carry it off.

We reexamine two different proposals for implementing the Penrose process. In one scheme, the black hole swallows the negative energy particle \cite{Misner}; in the other, the black hole {\it gradually} absorbs the particle's negative energy and angular momentum \cite{Christ}.

\section{Development}

Both approaches consider the particle to be  moving right at the  horizon in the equatorial plane of the black hole in the tangential direction,  i.e., it has the minimum energy $E$ consistent with its angular momentum $L$. For such a particle with mass $m$ there is a simple relation between  $L$ and  $E$ obtained by writing $p^2=g^{\mu \nu}p_{\mu} p_{\nu} =-m^2$ in terms of the Boyer-Lindquist coordinates \cite{Misner} \cite{Christ} \cite{Boyer};  when evaluated at $r=r_+$ it becomes

\begin{equation} 
L=F(M,J)E
 \label{eq:LFE}
\end{equation}
where $M$ and $J$ are the mass and angular momentum of the black hole, 

\begin{equation} 
F(M,J)=\frac{2M^2r_+}{J}=2MK(u) ,
 \label{eq:F(M,J)}
\end{equation}
and $r_+$ is the radius of the horizon,

\begin{equation}
r_+=M+\sqrt{M^2-J^2/M^2} = MuK(u).
 \label{eq:horizonJ}
\end{equation}
A useful relation  is 
 
 \begin{equation}
 r_+^2+J^2/M^2=2Mr_+=2M^2uK(u).
 \label{eq:relation}
 \end{equation}

We have introduced the dimensionless variable $u=J/M^2$ which ranges from $0$ to $1$, with zero representing a Schwarzschild black hole and unity an extreme Kerr black hole. The function 
$K(u)$ is defined to be

\begin{equation}
K(u)=\frac{1+\sqrt{1-u^2}}{u}.
 \label{eq:K(u)}
\end{equation}
We use units in which $G=1$ and $c=1$. 

\subsection{The particle is swallowed}

Reference \cite{Misner} considers the possibility that the black hole swallows the particle. The new mass and angular momentum of the black hole  would become
\begin{equation}
M'=M+E
\label{eq:newM}
\end{equation}
and
\begin{equation}
J'=J+L
\label{eq:newJ}
\end{equation}

We now consider the implications of the area theorem \cite{Hawking} for this process. It says that the surface area of a black hole cannot decrease, and we now  show that as a consequence of this theorem a particle having negative energy $E$ and angular momentum $L$ related to $E$ via Eq.(\ref{eq:LFE}) cannot be swallowed by a black hole. The formula for the area is \cite{Misner}

\begin{equation}
A=8\pi M r_+=8\pi M^2uK(u).
 \label{eq:area}
\end{equation}
The ratio of the new area $A'$ to the original area $A$ is 

\begin{equation}
\frac{A'}{A}=\frac{M'^2}{M^2}\frac{1+\sqrt{1-u'^2}}{1+\sqrt{1-u^2}}
\label{eq:A'ovA}
\end{equation}
with 

\begin{eqnarray}
u'
&=&
\frac{J'}{M'^2}  \nonumber \\
&=&
\frac{M^2u+2MK(u)E}{(M+E)^2} \nonumber \\
&=&
\frac{u+2K(u)y}{(1+y)^2},
\label{eq:u'}
\end{eqnarray}
where we have made use of Eq.(\ref{eq:LFE})    and  defined $y\equiv{E/M}$. The allowed range of $u'$ is also $0\leq u' \leq 1$. 

\begin{figure}[p]
\centering
\includegraphics[width=10cm]{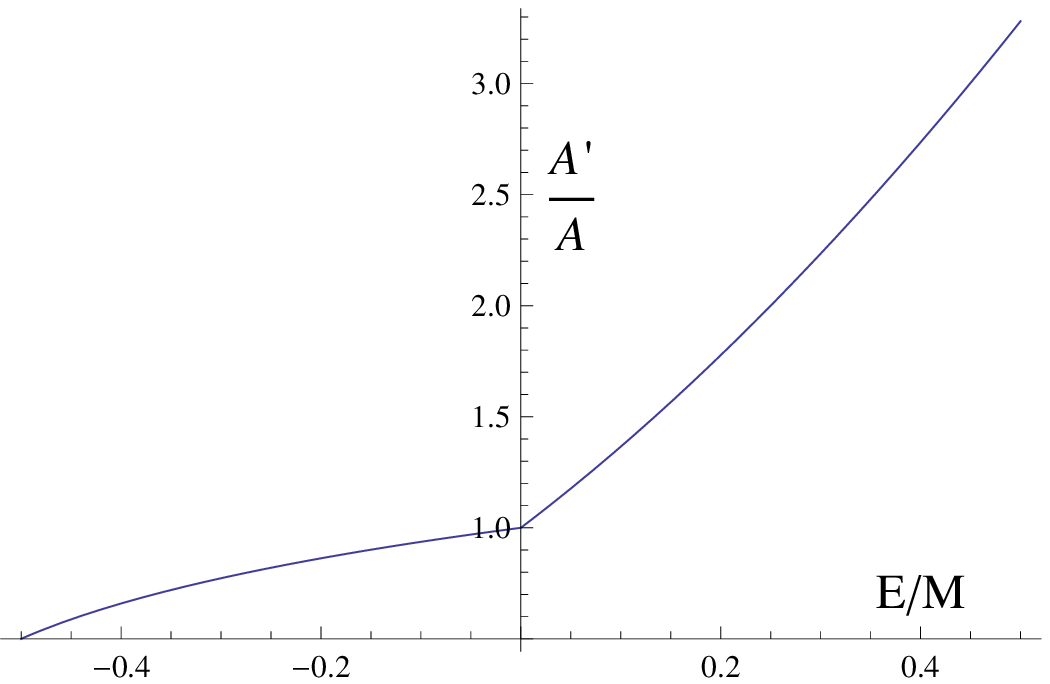}
\caption{Starting with an extreme Kerr black hole ($u=1$) of mass $M$,  $A'$ is its area after swallowing a particle at the horizon with energy E and no radial momentum, and $A$ is  its area before the absorption. The particle's angular momentum $L$ is given in Eq.(\ref{eq:LFE}).}{}
\label{Fig:AprovA}
\end{figure}

For the extreme Kerr case ($u=1$) there is a simple formula for   $A'/A$ obtained with a little algebra from Eqs.(\ref{eq:A'ovA}) and (\ref{eq:u'})

\begin{equation}
\frac{A'}{A}=(1+y)^2+ |y|\sqrt{2+4y+y^2}.
\label{eq:A'ovAext}
\end{equation}
This ratio is shown in Fig.\ref{Fig:AprovA}. For $y<0$ all the way down to $y=-1/2$, at which point $u'=0$, $A'/A<1$. This shows that if the particle moving at the horizon with no radial momentum has negative energy then the area of the black hole would decrease 
if it were to swallow the particle.  {\it This is forbidden by the area theorem} \cite{Hawking}. The same difficulty exists for a black hole  with initial angular momentum parameter value $u<1$. A particle with positive energy could be swallowed without violating the theorem, but serves no purpose insofar as the Penrose process is concerned. 

\subsubsection{Radial momentum}

If the particle at the horizon has some radial momentum $p^r$ as well as angular momentum then it becomes possible to swallow a particle with a limited amount of negative energy without violating the area theorem.  The relation between $L$, $E$, and $p^r$ again follows from $p^2=-m^2 $  and becomes  \cite{Misner}

\begin{equation}
L=2MK(u)E-MuK^2(u)|p^r|.
\end{equation}
Making use of this relation in the expression for $u'$ from Eq.(\ref{eq:u'}) and again specializing to the extreme Kerr case it can be shown that  there is a region of negative particle energy for which the area theorem would not be violated if the particle  is swallowed at the horizon. For small values of $|p^r|/M$ this region is approximately given by  $E/M>-\sqrt{|p^r|/2M}$.
 
[On a related matter it is pointed out in reference \cite{Misner} that if an extreme Kerr black hole swallows the particle the new angular momentum parameter $u'$ does not exceed unity, and consequently the horizon is not destroyed. But for black holes having $u=1-\epsilon$ arbitrarily close to the limiting value this is not necessarily the case. For small $\epsilon$ there is a range of positive energy values from $(\sqrt{2}-1)\sqrt{\epsilon}<E/M<(\sqrt{2}+1)\sqrt{\epsilon}$ for which $u'$ from Eq.(\ref{eq:u'}) is greater than unity.]

\subsection{Gradual absorption of the particle's energy}

Reference \cite{Christ} also considers the particle to be moving at the horizon with the minimum energy for a given angular momentum, i.e., they are related by Eq.(\ref{eq:LFE}).
In that reference \cite{Christ} it is proposed that instead of the black hole swallowing the particle a gradual exchange of energy and angular momentum takes place. Here we examine the consequences of treating the conservation of energy and angular momentum exactly, and show that it leads to very different conclusions from those found in \cite{Christ}. [That reference  used different symbols for $M, J$ and $L$.]  

For the gradual exchange mechanism we write the conservation laws in differential form

\begin{equation}
dM=-dE
 \label{eq:energycon}
\end{equation}
and 
\begin{equation}
dJ=-dL.
 \label{eq:angmomcon}
\end{equation}  
Imposing the two conservation laws on Eq.(\ref{eq:LFE}) gives

\begin{equation}
dJ=FdM-EdF
 \label{eq:cons}
\end{equation}
and using Eq.(\ref{eq:F(M,J)}) and the definition of $u$, Eq.(\ref{eq:cons})  becomes

\begin{equation}
2uMdM+M^2du=2K(M-E)dM-2ME\frac{dK}{du}du.
 \label{Eq:differentials}
\end{equation}
Gathering like terms together yields the coupled differential equations 

\begin{equation}
\frac{1}{M}\frac{dM}{du} =
\frac{1+2\frac{E}{M}\frac{dK}{du}}{2(K-u-K\frac{E}{M})} 
\label{eq:dMdu}
\end{equation}
and

\begin{equation}
\frac{dE}{du}=-\frac{dM}{du}
\label{eq:dEdu}
\end{equation}
Note that Eqs.(\ref{eq:dMdu}) and (\ref{eq:dEdu}) are invariant to the scale of $M$ provided  the initial condition on $E$ is altered by the same scale factor. Note also that $K\geq u$ and $dK/du$ is everywhere negative.

We can now see how the present development is related to that in \cite{Christ}.
  {\it If in}  Eq.(\ref{eq:dMdu})  {\it  the particle energy $E$ is set to  zero and required to remain zero throughout the process} it can be shown with some algebra to be equivalent to
 Eq.(5) in reference \cite{Christ}. [In \cite{Christ} that differential equation was obtained directly from Eq.(\ref{eq:LFE})  by setting $E=dM$ and $L=dJ$. This was said, without explanation, to represent conservation of energy and angular momentum.]

However,  setting $E\equiv 0$ is internally inconsistent as can be seen from  the analytic solution to the resulting differential equation,  given in reference \cite{Christ},  

\begin{equation}
2(\frac{M_{ir}}{M})^2=1+\sqrt{1-u^2}. 
\label{eq:Christeq}
\end{equation}
Here $M$ and $u$ are the initial values of the mass and angular momentum parameter of the black hole, and $M_{ir}$, called the "irreducible" mass, is the value that would be obtained if the black hole's angular momentum went to zero. In the case of an extreme Kerr black hole, it  would lead to its mass being reduced by 
$1-1/\sqrt{2}$= 29\% when its angular momentum goes to zero, i.e.,  when it becomes  a Schwarzschild  black hole.  This would mean, of course, that if the particle's energy were zero initially  it would  of necessity become 29\% of the black hole's mass by the end of the process, which is not negligible. [The inconsistency is even greater in the case of an extreme charged black hole, where 50\% of the black hole's mass is reduced \cite{Ruffini}.] We will show below that there are even more serious consequences of neglecting the terms involving the energy in 
Eqs.(\ref{eq:dMdu}) and (\ref{eq:dEdu}).

 Returning to the consequences of the area theorem,  and differentiating Eq.(\ref{eq:area}) gives
 
 \begin{equation}
 \frac{1}{8\pi M^2}\frac{dA}{du}=\frac{d(uK)}{du} + 2uK\frac{1}{M}\frac{dM}{du}
 \label{eq:cdAdu}
 \end{equation}
If in Eq.(\ref{eq:dMdu}) the particle energy is forced to be identically zero, as in reference \cite{Christ}, and the resulting expression for $dM/du$  is put into Eq.(\ref{eq:cdAdu}) it is straightforward to show that $dA/du$ would vanish. If this had been a valid procedure 
 Eq.(\ref{eq:Christeq}) could have been obtained directly from the condition that $A=constant$  without having to solve any differential equation.  It would have been a reversible process, and  also would have achieved the maximum possible  reduction in black hole mass. Unfortunately, it has led to the statement in the literature that  the Penrose process can be achieved in a manner that is arbitrarily close to being reversible. See, for example,  reference \cite{Hawking2}.

 What we now show, however, is that when energy and angular momentum conservation are imposed exactly, the procedure proposed in \cite{Christ}  cannot operate at all if $E$ and $L$ are negative because the evolution would violate the area theorem \cite{Hawking} by making the area {\it decrease}.  Inserting the complete expression for $dM/du$ from Eq.(\ref{eq:dMdu}) into (\ref{eq:cdAdu}) gives
  
 \begin{equation}
 \frac{dA}{du}=D \frac{K+2(K-u)\frac{dK}{du}}{(K-u-KE/M)(K-u)}\frac{E}{M}
  \label{eq:dAdu}
 \end{equation}
 where $D=8\pi M^2uK$, which is positive. The numerator of the first fraction in 
Eq.(\ref{eq:dAdu}) can be shown to be negative over the entire range of $u$; and {\it if $E$ is negative} the denominator is everywhere positive. Therefore, if $E$ is negative $dA/du$ is positive. This means that as the angular momentum and mass of the black hole decrease so would the area $A$, which is forbidden by the area theorem!  The conclusion is that the mechanism proposed in reference \cite{Christ}, consisting of a particle at the horizon moving tangentially and gradually giving up its negative energy and negative angular momentum to the black hole, cannot be valid.

There does  not appear to be any  reason, however, why the Penrose process of extracting energy and angular momentum from a black hole cannot proceed if the particle moving at the horizon  {\it initially} has zero or positive energy and corresponding angular momentum. It is seen from Eq.(\ref{eq:dAdu}) that if initially $E\geq 0$  and $E/M$ remains less than $(K-u)/K$ throughout the process, the area will increase as $u$ decreases, and hence is not forbidden. Before reaching that limit, however, the numerator of Eq.(\ref{eq:dMdu}) will vanish; that occurs when $E/M$ attains the value $1/(-2dK/du)$. At that point the mass of the black hole would increase as $u$ decreases further, effectively ending the process.

The sign results for $dM/du$ and $dA/du$ are summarized in the Table.
\begin{table}[h]
\centering
\begin{tabular}{|c||c|c|c|cl}
\hline
  &$E<0$&$E=0$&$0<E<\frac{M}{-2dK/du}$&$E=\frac{M}{-2dK/du}$ \\  
 \hline 
 $dM/du$&+&+&+&0 \\  
 \hline 
 $dA/du$&+&0&-&- \\  
 \hline
 \end{tabular}
\caption{The signs of the derivatives $dM/du$ and $dA/du$ in various regions of the particle energy $E$. For $0\leq{E}<{M/(-2dK/du)}$ the area theorem is not violated; and the black hole mass $M$ decreases  as the angular momentum parameter $u$ decreases.} {} 
\end{table}
 
We illustrate the possible removal of energy and angular momentum from the black hole with a numerical solution of  the coupled Eqs.(\ref{eq:dMdu}) and (\ref{eq:dEdu})  starting with the extreme Kerr case, and with the initial condition that the particle's energy $E$ and its angular momentum $L$ are both equal to zero. The mass scale is arbitrary  and we have taken $M(u=1)=1$. Due to the square root singularity in the function $K(u)$ at $u=1$ it was necessary to take the initial value of $u=u_0$ to be slightly less than unity, and let $u_0$ increase towards unity until the solution converged to sufficient accuracy.  

\begin{figure}[p]
\centering
\includegraphics[width=10cm]{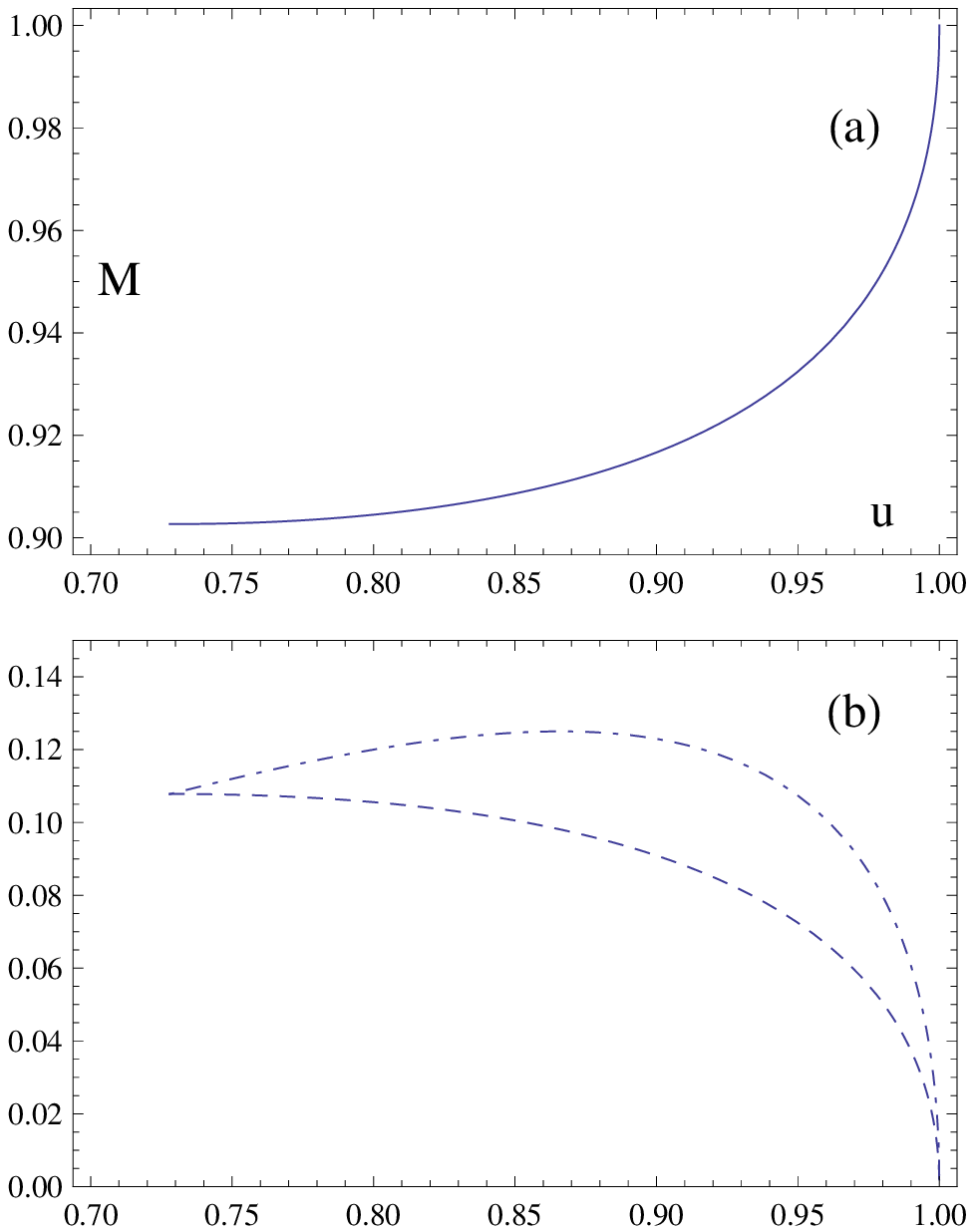}
\caption{Starting with an extreme Kerr black hole with angular momentum parameter        
$u=J/M^2=1$, and initial particle energy and angular momentum both equal to zero, the numerical solution of the coupled differential 
Eqs.(\ref{eq:dMdu}) and (\ref{eq:dEdu}) is shown. This implements the Christodoulou \cite{Christ} 
mechanism with energy and angular momentum conservation imposed exactly. (a) The black hole mass decreases by 9.7\% at $u=0.728$, and does not decrease any further. (b) The dashed curve is the ratio of the particle energy $E$ over the black hole mass $M$; the dot-dashed curve is the quantity $-1/(2dK/du)$.  It is shown in the text that when they become equal, $dM/du=0$.} {}
\label{Fig:graphcolumn}
\end{figure}

The result is shown in Fig.\ref{Fig:graphcolumn}. The black hole's mass and angular momentum both decrease until  $dM/du=0$ at $u=0.728$,  at which point $M=0.903$. This represents the maximum reduction in the black hole mass possible starting with an extreme Kerr black hole and using the mechanism in reference \cite{Christ}. At the end of the process $E=1-0.903$ and $E/M=0.107$. As expected from the discussion above and shown in part (b) of the figure, this value of $E/M$ is equal to $-1/(2dK/du)$ evaluated at $u=0.728$, the place at which the numerator of Eq.(\ref{eq:dMdu}) vanishes. If the initial value of $E$ were positive rather than zero the process would end at an even larger value of $M$.

[Starting with the black hole having $u=0.728$, the same mechanism could  operate with a second particle. If it has zero energy initially the black hole mass and angular momentum would decrease further, but its mass decreases by only another 3.9\%, ending at $u=0.42$.]

\section{Conclusion}

We have examined the transfer of energy  and angular momentum from a rotating black hole to a particle moving  in the equatorial plane right at the horizon, in two different scenarios.  
In the first \cite{Misner} the black hole swallows the particle. If the particle has negative energy, and the minimum amount  consistent with its angular momentum, then we have shown that the process cannot take place because it would violate the requirement that the the black hole area not decrease. If, on the other hand, the particle has some radial momentum in addition to the angular momentum, then the process can proceed; for a small amount of radial momentum the requirement is that $E/M>-\sqrt{|p^r|/2M}$.

In the other  scenario the particle and black hole gradually exchange energy and angular momentum \cite{Christ}. By imposing energy and angular momentum exactly we arrive at very different results from those in reference \cite{Christ}.  In that paper it appeared that the black hole could give up all its angular momentum to the particle, thereby becoming a Schwarzschild black hole; and its mass would decrease by the  theoretical maximum of $1-1/\sqrt{2}=29$\%. 
We find instead two conclusions. First, that {\it the mechanism proposed in reference \cite{Christ} cannot operate  at all if the particle has negative energy and angular momentum because it would violate the area theorem} \cite{Hawking}. 

The second result is that the same mechanism  can operate if the particle has {\it positive} energy and angular momentum, but nevertheless extracts energy and angular momentum from the black hole. It falls far short of attaining the theoretical maximum, however. Starting with an extreme Kerr black hole and a particle initially having zero energy and angular momentum, the black hole's mass decreases by 9.7\% when its angular momentum parameter $u$ attains the value $u=0.728$. At that point the ratio of the particle's energy to the black hole mass is $E/M=0.107$, and $dM/du$ reaches a minimum. To reduce the black hole's mass even further would require  repeating the process with a second particle having zero or positive energy.

\end{document}